\documentclass[aps,prb,showpacs,floatfix,amsmath,amssymb,superscriptaddress,reprint]{revtex4-2}
\usepackage{tikz}
\usetikzlibrary{shapes.geometric, arrows.meta, positioning}
\usepackage{afterpage} 
\usepackage{float}
\usepackage{tikz}
\usepackage{bbold}
\usepackage[normalem]{ulem}
\usepackage{color}
\usepackage{graphicx}
\usepackage{epsfig}
\usepackage{setspace}
\usepackage{tikz}
\usepackage{tikz-feynman}
\tikzfeynmanset{compat=1.1.0}
\usepackage{amsmath}
\usepackage{amssymb,amsfonts,mathrsfs,mathtools,braket}
\usepackage{verbatim}
\usepackage{bibentry}

\usepackage{cancel}
\usepackage[percent]{overpic}
\usepackage{xcolor}

\newcommand{\bea}{\begin{eqnarray}}
\newcommand{\eea}{\end{eqnarray}}

\usepackage[pdfencoding=auto, psdextra, colorlinks=true, linkcolor=blue,citecolor=blue, urlcolor=blue]{hyperref}

\begin{document}

\author{Pragya Chaudhary}
\email{p20240079@pilani.bits-pilani.ac.in}
\affiliation{Department of Physics, Birla Institute of technology and Science, Pilani 333031, India}

\title{Reinterpreting Memory Effects in Nonequilibrium Systems: From Temporal Dynamics to Steady-State Signatures via NEGF}

\begin{abstract}
 \textbf{ABSTRACT}:We investigate memory effects and quantum transport in two-dimensional lattice systems within the framework of non-equilibrium Green's functions and Schwinger-Keldysh non-equilibrium quantum field theory. Starting from a 2D tight-binding Hamiltonian, we employ the Dyson expansion on the Keldysh contour and the second-order Born and self-consistent Born Approximation to derive the electronic self-energies associated with elastic and inelastic scattering mechanisms.Static disorder produces a local self-energy and a rapidly decaying memory kernel, characteristic of Markovian dynamics, whereas electron-phonon coupling generates temporally nonlocal self-energies and genuine Non-Markovian behavior. We demonstrate that these distinct memory signatures are directly reflected in the spectral function, which we propose as a diagnostic probe of non-equilibrium memory effects. Further we explore 1PI and 2PI effective actions to see their memory perspectives studying their coarse-graining behavior. Building on this theoretical framework, we further apply the conventional NEGF formalism to two paradigmatic two-dimensional models-the Hofstadter and an RKKY-coupled system to explore how different microscopic Hamiltonians influence Markovian and Non-Markovian nature. Our results provide a unified connection between scattering mechanisms, memory effects, and quantum transport in low-dimensional systems.
\end{abstract}
\maketitle
\section{Introduction} 
Quantum transport in low-dimensional and nanoscale systems \cite{buot2009nonequilibrium} is fundamentally influenced by interactions, coherence and coupling to external environments, rendering nonequilibrium effects central to both theory and experiment\cite{Datta_2005}. While many transport approaches rely on Markovian approximation\cite{merkli2022dynamics} assuming that the system instantaneously loses information about its past, It is now well understood that this assumption can fail in interacting or energy-exchange-dominated regimes, giving rise to non-Markovian dynamics\cite{Breuer,shrikant2023quantum} with explicit memory dependence .Within nonequilibrium quantum field theory\cite{Berges_2004}, memory effects arise naturally from the Dyson equation formulated on the Schwinger-Keldysh contour\cite{leonardi1986dicke,kamenev2023field,sieberer2016keldysh}, whose real time projection leads to the Kadanoff-Baym equations for two-time Green's functions \cite{HaugHJauho,schwinger1961brownian,keldysh2024diagram,kadanoff2018quantum}.In nonequilibrium quantum transport, coupling electronic degrees of freedom to a phonon bath naturally introduces dissipative and memory effects, making electron--phonon systems an ideal framework for investigating the emergence of Markovian and non-Markovian behavior\cite{morozov2000non,PhysRevLett.104.157401,iotti2020energy}.
In this context, recent developments have emphasized semiclassical and hybrid NEGF approaches\cite{Datta_2005} that retain the temporal structure of electron-phonon self-energies while remaining computationally tractable, demonstrating how phonon dynamics on timescale comparable to electronic transport leads to nonlocal corrections beyond standard Markovian treatments \cite{ochoa2025semiclassical}.The convolution structure of these equations introduces self-energy kernels that are non-local in time, thereby encoding the system's dynamical history. When the self-energy is sharply localized in time, the resulting dynamics become effectively Markovian\cite{jeknic2016dynamical}, whereas extended temporal Kernels lead to long-lived correlations and non-Markovian evolution\cite{shrikant2023quantum}. This framework has been widely employed to analyze interacting transport, dissipation, and relaxation processes in mesoscopic and nanoscale systems\cite{Bonitz,Datta_2005,stefanucci2013nonequilibrium}.Recent studies have established that non-Markovian memory effects in open quantum systems\cite{PhysRevLett.109.170402,breuer2016colloquium,de2017dynamics,bi2026generalized} can be systematically understood through frequency-domain formulations that retain the full time-nonlocal structure\cite{Review}. By analyzing memory kernels and reduced propagators in the frequency domain, it has been shown that deviations from Markovian dynamics are encoded in nontrivial spectral features arising from finite bath correlation times\cite{mikhailov2021non} and structured environmental couplings\cite{ivander2024unified}. Complementary approaches demonstrate that non-Markovian behavior can be directly diagnosed from steady-state spectral signatures\cite{li2025spectral},such as peak splitting and frequency dependent broadening,without relying on explicit time-resolved dynamics\cite{keefe2025quantifying}. Together with frequency-domain master equation\cite{dann2018time,chiang2021non,debecker2024spectral} formulations that connect bath spectral densities to observable line shapes,these works provide a unified framework connecting memory kernels\cite{kleinekathofer2004non} to observable frequency-dependent response signatures of non-Markovian behaviour beyond standard Markovian approximations.

From an experimental perspective, elastic and inelastic scattering provide a natural framework for distinguishing qualitatively different nonequilibrium transport regimes. Elastic scattering, arising from static disorder such as impurities, conserves electronic energy and primarily leads to phase decoherence without energy relaxation\cite{Datta_2005,HaugHJauho,mahan2013many,rammer2007quantum}. In contrast, inelastic scattering most notably due to electron-phonon interactions that involves energy exchange with the environment and generates self-energies that are non-local in time\cite{breuer2002theory}. Within the second order Born Approximation\cite{schlunzen2020ultrafast}, such processes mix electronic states separated by the phonon energy, producing self-energy kernel with a finite temporal extent \cite{HaugHJauho,mahan2013many}. These effects are central to transport experiments in molecular junctions, low-temperature semiconductor systems, and time-resolved spectroscopies, where energy relaxation and persistent temporal correlations have been observed \cite{SentefMichael,Eckstein_2010,PhysRevLett.68.2}. Consequently, understanding how elastic and inelastic scattering imprint distinct temporal signatures on nonequilibrium Green's functions is essential for connecting microscopic dynamics to measurable transport properties.\\
In this work, we investigate the microscopic origin and physical consequences of Markovian and non-Markovian behavior in quantum transport systems. Section \ref{second}, reviews the formal foundation
of memory effects through the Kadanoff–Baym equations, emphasizing on the role of temporal convolution and the emergence of nonlocal-in-time dynamics. In Section \ref{Third}, we analyze Markovian and
non-Markovian signatures using the spectral function, contrasting elastic and inelastic scattering processes. Section \ref{Fourth} further examines these signatures by comparing the Born and self-consistent Born approximations, highlighting how different levels of approximation influence dynamical memory effects. In Section \ref{Fifth}, we present a renormalization group(RG) interpretation through particle irreducible effective action to provide a scaling perspective on the transition between Markovian and non-Markovian regimes. Finally, Section \ref{Sixth} clarifies the dynamical structure and flow of memory effects in interacting quantum systems via conventional NEGF formalism\cite{KeremCamsari}.
\section{Memory Effects in Quantum Transport:Formalism and Microscopic Origins}\label{second}
\subsection{Contour Ordered Green's Functions and Dyson Equation}
The starting point of nonequilibrium quantum transport theory is the contour-ordered Green's function defined on the Schwinger-Keldysh closed time contour C \cite{schwinger1961brownian,keldysh2024diagram,Schwinger}
\begin{equation}
G(t,t') = -i \langle T_C \psi(t) \psi^\dagger(t') \rangle
\end{equation}

The exact dynamics are governed by the Dyson equation on the contour,
\begin{equation}
    G=G_0+G_0\otimes \Sigma \otimes G
\end{equation}

Where $\Sigma$ is the interaction self-energy and $\otimes$ denotes convolution \cite{keldysh2024diagram,stefanucci2013nonequilibrium}.
\subsection{Kadanoff-Baym Equation and Temporal Convolution}
Projection of the Dyson equation onto real-time branches yields the Kadanoff-Baym equations \ref{A}(Appendix [A]) in two-time Green's function \cite{kadanoff2018quantum,DANIELEWICZ1984239,HaugHJauho}. For the retarded component one obtains, 
\begin{equation}
i\frac{\partial}{\partial t} G^r(t, t')=h(t)G^r(t, t')+\int_{t_0}^{t} d\bar{t} \,\Sigma^r(t, \bar{t}) G^r(\bar{t}, t')\label{eq3}
\end{equation}
the corresponding equation holds for derivatives with respect to t. Here h(t) is the single-particle Hamiltonian and $\Sigma(t, t')$ incorporates interactions and environment effects.
The time-convolution integral over $\bar{t}$ couples the present dynamics to all earlier times. This convolution structure is the precise mathematical origin of temporal correlation in non-equilibrium transport and is universally present in interacting systems \cite{Bonitz,stefanucci2013nonequilibrium}.
\subsection{Elastic Scattering and Time-Local Self-Energies}
Elastic scattering processes conserve electronic energy and arise from static disorder, impurities, or interface roughness. Within the NEGF framework, such processes are commonly modeled by self-energies that are local in time,
\begin{equation}
    \Sigma_{el}^r(t, t') \propto -i \Gamma \delta(t-t'),
\end{equation}
which reduces the Kadanoff-Baym equation to a time-local form \cite{Datta_2005}. In this limit, the Green's function evolution depends only on it's instantaneous value, and temporal correlations decay on timescales set by the elastic scattering rate. This regime serves as a natural baseline for identifying deviations induced by interaction-driven temporal nonlocality.
\subsection{Electron-Phonon Coupling as a Controlled Source of inelastic Scattering}
Coupling an electronic system to a phonon bath provides a controlled and physically transparent description of inelastic scattering, enabling energy exchange and dissipation through environmental degrees of freedom. A widely used model for electron-phonon interaction is the linear and corresponding Hamiltonian is given as,
\begin{equation}
    H_{e-ph}= \sum_{i,q} g_q c_{i}^{\dagger} c_{i}(b_q +b_{q}^{\dagger})
\end{equation}
where $c_{i}^{\dagger} (c_{i})$ and $b_{q}^{\dagger}(b_{q})$ are electronic and phononic creation(annihilation) operators respectively, and $g_{q}$ denotes the coupling strength \cite{mahan2013many,HaugHJauho}. Owing to their broad frequency spectrum, phonons provide an efficient microscopic mechanism for energy relaxation in nanoscale conductors \cite{Datta_2005}.
Within the second-order Born approximation, the electron-phonon self-energy factorizes into electronic and phononic Green's functions. In the time domain, the retarded component takes the form
\begin{equation}
    \Sigma_{e\text{--}ph}^r(t, t')= \sum_{q}|g_q|^2 G(r,r')D_{q}^r(t-t'),
\end{equation}
where $D_{q}^r(t-t')$ is the retarded phonon Green's function \cite{mahan2013many,stefanucci2013nonequilibrium}. The finite decay time of $D_{q}(t)$, set by the phonon spectrum and damping, determines the temporal extent of the self-energy kernel. As a result, electronic dynamics depend explicitly on their past evolution, providing a microscopic origin of interaction-induced temporal correlations within the Kandanoff-Baym framework \cite{kadanoff2018quantum,DANIELEWICZ1984239,HaugHJauho}.
\subsection{Role of This Framework}
The mathematical structure summarized above is well established in nonequilibrium quantum field theory and forms the foundation for analyzing temporal correlations in interacting transport systems. In the following section, this framework is employed to explicitly evaluate interaction-induced temporal signatures in concrete lattice models, thereby connecting the general formalism to quantitative numerical results.

\section{Markovian and Non-Markovian Signature via Spectral Function For Elastic and Inelastic Scattering}\label{Third}
In this section, we explore that how different microscopic scattering mechanisms manifest as distinct temporal signatures in the spectral function\cite{mahan2013many}, focusing on elastic and inelastic processes within a two-dimensional tight-binding framework. Starting from a common electronic Hamiltonian, we incorporate scattering through appropriate self-energy terms and study the resulting spectral function in the time domain, which serves as a direct probe of temporal correlations in non-equilibrium transport.
\subsection{Two-Dimensional Tight-Binding Hamiltonian}
We consider a generic two-dimensional square-lattice model describing non-interacting electrons,
\begin{equation}\label{eq7}
    H_{0}=\sum_{i,j}t_{ij}c_{i}^{\dagger}c_{j}+\sum_{i}U_{i}c_{i}^{\dagger}c_{i}
\end{equation}

Here $c_{i}^{\dagger}$ and $c_{i}$ creates and annihilates an electron at lattice site i respectively and $t_{ij}$ denotes nearest-nearest neighbor hopping amplitudes, and $U_{i}$ is the on-site potential. Different choices of $t_{ij}$ and $U_{i}$ allows us to explore for different models, such as Hofstadter Hamiltonian in a magnetic field or long-range hopping models motivated by RKKY-type interactions, which are explored in the next part of our work.
The effect of scattering is incorporated through self-energies in the nonequilibrium Green's function formalism. The retarded Green's function satisfies equation \ref{eq3}.

where $\Sigma^r(t, \bar{t})$ is the retarded self-energy describing coupling to disorder, phonons or other electrons.
The time-domain spectral function is defined as 
\begin{equation}
    A(\tau)= -2\,\mathrm{Im}\,\mathrm{Tr}\, G^r(\tau),   \qquad \tau=t-t'.
\end{equation}
The decay profile of $A(\tau)$ provides direct information about the temporal structure of the self-energy kernel and hence about the nature of the scattering process.
 \subsection{Elastic Scattering: Time-Local Self-Energy and Markovian Dynamics}
 Elastic scattering arises from static disorder, impurities, or interface roughness and does not involve energy exchange. Microscopically, this corresponds to a potential
 \begin{equation}
     H_{el}=\sum_{i}U_{i}c_{i}^{\dagger}c_{i}
 \end{equation}
where $U_i$ represents the impurity potential on-site.
The $U_i$ are treated as random variables describing static disorder.

Physical observables are obtained by averaging over disorder realizations.
We decompose the random potential as
\begin{equation}
U_i = U + \delta U_i ,
\qquad
\langle \delta U_i \rangle = 0 ,
\end{equation}
where $U$ represents a uniform background shift, while $\delta U_i$ describes
fluctuations about the mean.
Assuming uncorrelated (white-noise) disorder, the second moment is taken as
\begin{equation}
\langle \delta U_i \delta U_j \rangle
=
2\Gamma\,\delta_{ij},
\end{equation}
where $\Gamma$ is the elastic scattering rate.
This assumption reflects the physical picture of randomly distributed,
uncorrelated static impurities. 
 By the Born approximation, the elastic self-energy is given by
\begin{equation}
\Sigma_{\text{el}}(i,j;t,t')
=
\langle \delta U_i \delta U_j \rangle\,
G(i,j;t,t').
\end{equation}
Using the disorder correlator above, the self-energy becomes local in space
and time,
\begin{equation}
\Sigma_{\text{el}}(t,t')
\propto
\delta(t-t').
\end{equation}

Fourier transforming to frequency space yields a frequency-independent
retarded self-energy,
\begin{equation}
\Sigma^R_{\text{el}}(\omega)
=
U - i\Gamma ,
\end{equation}
where the real part $U$ produces an energy shift, while the imaginary part
$\Gamma$ corresponds to a finite elastic lifetime, consistent with Fermi's
golden rule.
Transforming back to the time domain, the retarded self-energy for $\tau>0$
can be written as
\begin{equation}
\Sigma^R_{\text{el}}(\tau)
=
-\,i\Gamma\, e^{-\Gamma\tau},
\end{equation}
which represents a rapidly decaying, short-lived response.
In the limit $\Gamma \to \infty$, this expression approaches a delta function,
$\Sigma^R_{\text{el}}(\tau)\propto\delta(\tau)$.

 Now the memory kernel is defined as
\begin{equation}
A_{el}(\tau)
=
-2\,\mathrm{Im}\,\Sigma_{el}^R(\tau).
\end{equation}
Substituting Eq.~(16
) yields
\begin{equation}
A_{\text{el}}(\tau)
=
2\Gamma\, e^{-\Gamma\tau}.
\end{equation}

This kernel is sharply peaked at $\tau=0$ and decays on a timescale
$1/\Gamma$, indicating that elastic scattering induces only short-lived memory (Markovian effects).
\begin{figure} [h] 
    \centering
    \includegraphics[width=1.0\linewidth]{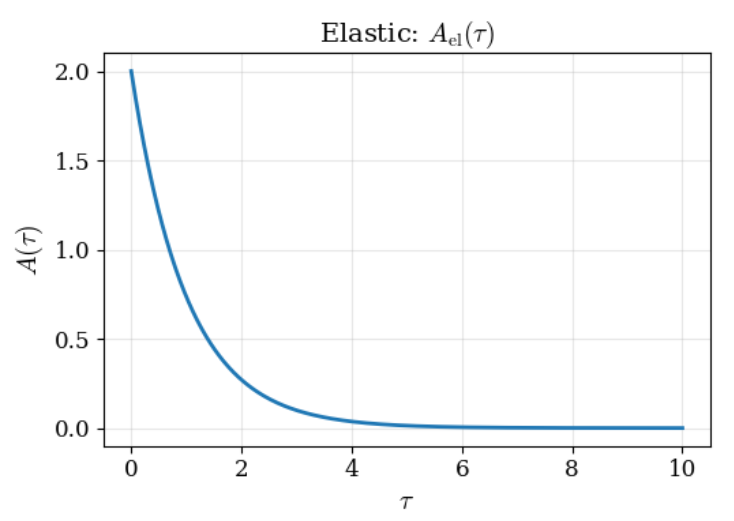}
    \caption{ Decaying nature corresponds to the elastic scattering}
\end{figure}
\subsection{Inelastic Scattering: Electron-Phonon Coupling and Temporal Non-Locality}
Inelastic scattering refers to processes in which the electronic subsystem exchanges energy with additional degrees of freedom. So,  coupling the electronic system to a phonon bath provides a controlled description of inelastic scattering.
The electron--phonon interaction \cite{HaugHJauho,ochoa2025semiclassical},
\begin{equation}
H_{e\text{-}ph} = \sum_{i,q} g_q\, c_i^\dagger c_i \left(b_q + b_q^\dagger\right),
\end{equation}
introduces a set of bosonic modes that act as an effective environment for the electrons.
Phonons possess a broad spectrum of frequencies and can absorb or emit energy, allowing for irreversible relaxation processes.

Within the Born approximation \ref{B}(Appendix [B]), the resulting retarded self-energy factorizes into an electronic and a phononic contribution \cite{HaugHJauho},
\begin{equation}
\Sigma^R_{e\text{-}ph}(t,t') \propto g_q^2\, G_0(t,t') D^R_q(t-t'),
\end{equation}
where $D^R_q$ denotes the retarded phonon Green's function.
The time dependence of $D^R_q$ directly determines the memory kernel entering the Kadanoff--Baym equations.

Because phonons have finite lifetimes, their retarded Green's functions decay slowly in time, leading to oscillatory, long-ranged memory kernels.
This produces intrinsically non-Markovian dynamics and allows for a clear distinction between elastic scattering, which does not exchange energy, and inelastic scattering mediated by phonon emission and absorption.
Again, For a harmonic phonon mode $q$ with frequency $\omega_q$ and damping rate $\gamma_q$,
the retarded phonon Green's function is\cite{mahan2013many}
\begin{equation}
D_q^R(\tau)
=
-\,\theta(\tau)\,
e^{-\gamma_q \tau}
\sin(\omega_q \tau),
\end{equation}
where $\tau = t - t'$.

Substituting this expression yields the retarded inelastic self-energy
\begin{equation}
\Sigma^R_{\text{inel}}(\tau)
=
-\,\theta(\tau)
\sum_q |g_q|^2
e^{-\gamma_q \tau}
\sin(\omega_q \tau).
\end{equation}
The memory kernel is defined as the imaginary part of the retarded self-energy,
\begin{equation}
A(\tau)
=
-\,2\,\mathrm{Im}\,\Sigma_{inel}^R(\tau).
\end{equation}

Using Eq.~(32), the inelastic memory kernel becomes
\begin{equation}
A_{\text{inel}}(\tau)
=
2\sum_q |g_q|^2
e^{-\gamma_q \tau}
\sin(\omega_q \tau).
\end{equation}

\begin{figure}
    \centering
    \includegraphics[width=1.0\linewidth]{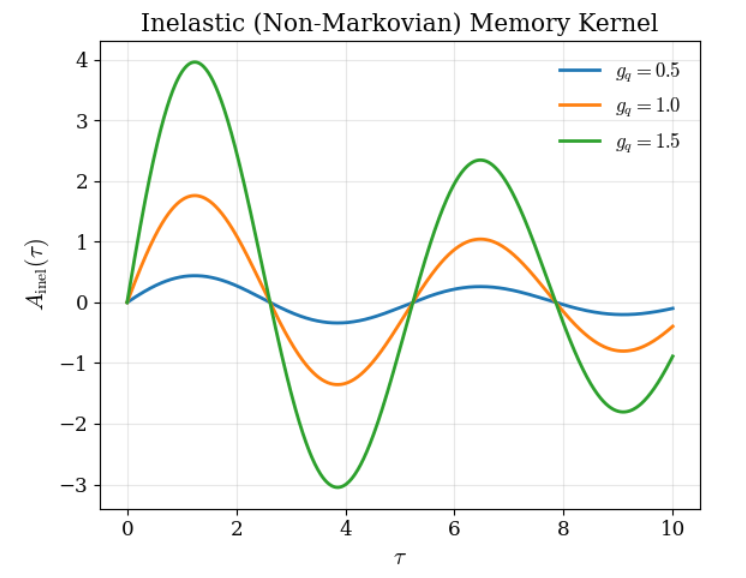}
    \caption{Oscillatory nature corresponds to inelastic Born approximation for varying coupling constants $g_q$.}\label{inelas}
\end{figure}

This oscillatory and slowly decaying kernel reflects the long-lived memory \cite{HaugHJauho}(Non-Markovian)effects associated with inelastic scattering via phonon emission and absorption as shown in Fig.\ref{inelas}.\\

\subsection{Consistency of the Markovian Limit and Identification of the Elastic Rate}
In the Markovian limit, inelastic scattering mediated by a phonon
bath reduces to an effective elastic scattering process. In this subsection, we
justify the identification of the elastic scattering rate
and demonstrate its consistency at the level of the memory kernel and the
equations of motion.The inelastic memory kernel entering the Kadanoff--Baym equations is given by
\begin{equation}
A_{\mathrm{inel}}(\tau)
=
2|g_q|^2 e^{-\gamma_q\tau}\sin(\omega_q\tau).
\label{eq:ainel_exact}
\end{equation}
In the fast-bath limit $\gamma \gg \omega_q$, the bath correlations decay on a
timescale much shorter than the intrinsic system dynamics. Consequently, the
oscillatory factor may be expanded for small $\tau$,
\begin{equation}
\sin(\omega_q\tau) \simeq \omega_q\tau,
\end{equation}
leading to
\begin{equation}
A_{\mathrm{inel}}(\tau)
\approx
2|g_q|^2\,\omega_q\,\tau\,e^{-\gamma\tau}.
\label{eq:ainel_fastbath}
\end{equation}
The memory kernel enters the equations of motion only through the convolution
\begin{equation}
I(t)
=
\int_0^\infty d\tau\;
A_{\mathrm{inel}}(\tau)\,F(t-\tau),
\label{eq:memory_convolution}
\end{equation}
where $F(t)$ denotes a slowly varying system quantity (e.g., a Green's function
or density matrix element).

Substituting Eq.~\eqref{eq:ainel_fastbath} into
Eq.~\eqref{eq:memory_convolution} yields
\begin{equation}
I(t)
=
2|g_q|^2\omega_q
\int_0^\infty d\tau\;
\tau e^{-\gamma\tau}
F(t-\tau).
\end{equation}

Introducing the rescaled variable $u=\gamma\tau$,
with $d\tau = du/\gamma$, one finds
\begin{equation}
I(t)
=
2|g_q|^2\omega_q
\int_0^\infty du\;
\frac{u}{\gamma^2} e^{-u}
F\!\left(t-\frac{u}{\gamma}\right).
\end{equation}

Since $F(t)$ varies on timescales much longer than $1/\gamma$, it may be expanded
as
\begin{equation}
F\!\left(t-\frac{u}{\gamma}\right)
=
F(t) + \mathcal{O}\!\left(\frac{1}{\gamma}\right).
\end{equation}
Retaining the leading term gives
\begin{equation}
I(t)
=
2\frac{|g_q|^2\omega_q}{\gamma^2}
F(t)
\int_0^\infty du\; u e^{-u}.
\end{equation}
Using $\int_0^\infty du\,u e^{-u}=1$, we obtain
\begin{equation}
I(t)
=
2\frac{|g_q|^2\omega_q}{\gamma^2}
F(t).
\label{eq:It_result}
\end{equation}
For comparison, a purely elastic (Markovian) memory kernel has the form
\begin{equation}
A_{\mathrm{el}}(\tau)
=
2\Gamma\,\delta(\tau),
\end{equation}
which yields
\begin{equation}
\int_0^\infty d\tau\;
A_{\mathrm{el}}(\tau)\,F(t-\tau)
=
2\Gamma\,F(t).
\end{equation}
Matching this expression with Eq.~\eqref{eq:It_result} uniquely fixes the elastic
scattering rate as
\begin{equation}
\Gamma = \frac{|g_q|^2\,\omega_q}{\gamma^2}.
\end{equation}
The above identification reflects the fact that, in the fast-bath\cite{Fast-Bath} 
limit, phonon-induced scattering becomes effectively instantaneous. The bath
correlation time $1/\gamma$ is much shorter than all system timescales, causing
the nonlocal memory kernel to collapse into a local-in-time contribution. As a
result, inelastic scattering reduces to elastic scattering with rate $\Gamma$,
demonstrating that elastic dynamics is the Markovian limit of inelastic
electron--phonon interactions.
\section{BORN VERSUS SELF-CONSISTENT BORN APPROXIMATION: ELASTIC AND
INELASTIC SCATTERING}\label{Fourth}
In this section we compare the lowest-order Born approximation with the self-consistent Born approximation(SCBA) within the nonequilibrium Green's function(NEGF) framework. The comparison is carried out for both elastic(static disorder) and inelastic(electron-phonon) scattering, with particular emphasis on the temporal structure of the resulting self-energies. We show that while elastic scattering leads to a strictly local (Markovian) self-energy irrespective of self-consistency, inelastic scattering generates a finite temporal memory kernel whose strength and shape are renormalized by self-consistency.\\\
Within NEGF theory, interactions and coupling to external degrees of freedom are incorporated through the self-energy $\Sigma$. The retarded Green's function satisfies the Dyson equation
\begin{equation}
    G^R=G_{0}^R+G_{0}^R\Sigma^RG^R+.......
\end{equation}
where $G_{0}^R$ denotes the non-interacting(bare) Green's function and $\Sigma^R$ encodes the effects of scattering processes.\\
To second order in the interaction strength $g_q$, the self-energy takes the generic form 
\begin{equation}
    \Sigma(t,t')=ig_q^2G(t,t')D(t,t'),
\end{equation}
where G is the electronic Green's function and D is the propagator of the bath degrees of freedom(e.g., disorder or phonons). Two commonly used approximations follow from different choices of the Green's function entering this expression:\\
\textbf{Born approximation:} $G\rightarrow G_{0}$, yielding $\Sigma_{Born}(1,2)=ig_q^2G_{0}(t,t')=ig_q^2G_{0}(t,t')D(t,t')$\\

\textbf{Self-consistent(SCBA):} $G\to G$,  yielding $\Sigma_{Born}(1,2)=ig_q^2G(t,t')=ig_q^2G(t,t')D(t,t')$
\subsection{Elastic scattering(static disorder)}
We first consider elastic scattering due to static disorder, described by the Hamiltonian 
\begin{equation}
    H_{el}=\sum_{i}U_{i}c_{i}^{\dagger}c_{i}
\end{equation}
where $U_{i}$ is a random potential. The disorder is assumed to be uncorrelated and Gaussian, with \\
\begin{equation}
\langle U_{i}U{i}\rangle=U^2 \delta_{ij}
\end{equation}
Since the potential is static, the bath correlator is instantaneous:
\begin{equation}
    D(t)=\delta(t)
\end{equation}
Now, using the instantaneous correlator in the Born self-energy yields 
\begin{equation}
    \Sigma_{Born}=U^2G_{0}(t)\delta(t)=U^2\delta(t)
\end{equation}
which is strictly local in time. Fourier transformation gives a purely imaginary, energy-independent retarded self-energy.
\begin{equation}
\Sigma_{Born}^R=-i\Gamma,  \quad  
\Gamma=\pi U^2\rho_{0}
\end{equation}
where $\rho_{0}$ is the density of the states of the clean system. The imaginary part $\Gamma$ corresponds to a finite elastic scattering rate and produces a constant lifetime broadening of the electronic states.\\
Now, within Self-Consistent Born Approximation the self energy is given by
\begin{equation}
    \Sigma_{Born}=U^2G(t)\delta(t)=U^2\delta(t).
\end{equation}
Since the bath correlator is instantaneous, the temporal structure of the self-energy remains local even after self-consistency is imposed. As a result,
\begin{equation}
    \Sigma_{SCBA}=\Sigma_{Born}.
\end{equation}
Thus, elastic scattering from static disorder generates a purely Markovian kernel, and self-consistency does not modify it's time dependence. While SCBA can renormalize spectral properties through feedback in frequency space, it does not introduce temporal memory in this case. 
\subsection{Inelastic scattering(electron-phonon interaction)}
We now turn to inelastic scattering arising from electron-phonon coupling, described by 
\begin{equation}
    H_{e-ph}=\sum_{i}g_{q}c_{i}^\dagger c_{i}(b_{q}+b_{q}^\dagger).
\end{equation}
Unlike static disorder, phonons are dynamical degrees of freedom. The retarded phonon propagator is taken as 
\begin{equation}
    D^{R}(t)
= \theta(t) e^{-\gamma t} \sin(\omega_0 t),
\end{equation}
which possesses a finite correlation time $1/\gamma$.
Within SCBA the electronic propagator acquires a lifetime
\begin{equation}
G^{R}(\omega)
= \frac{1}{\omega-\epsilon+i\Gamma},
\qquad
\Gamma \sim g_q^2 \rho_0.
\end{equation}

Fourier transforming,
\begin{equation}
G^{R}(t)
= -i \theta(t) e^{-i \epsilon t} e^{-\Gamma t}.
\end{equation}

Substituting into the self-energy,
\begin{align}
\Sigma_{\text{SCBA}}^{R}(t)
&\propto g_q^2 G^{R}(t) D(t) \\
&\propto g_q^2 e^{-\Gamma t} e^{-\gamma_{q} t} \sin(\omega_q t).
\end{align}

Therefore
\begin{equation}
\Sigma_{\text{SCBA}}^{R}(t)
\propto
e^{-(\gamma_{q}+\Gamma)t} \sin(\omega_q t)
\end{equation}
The memory kernel is defined as the imaginary part of the retarded self-energy,
\begin{equation}
A(\tau)
=
-\,2\,\mathrm{Im}\,\Sigma_{inel}^R(\tau).
\end{equation}
Using Eq.~(23), the inelastic memory kernel becomes
\begin{equation}
A_{\text{inel}}(\tau)
=
2\sum_q |g_q|^2
e^{-(\gamma_q+\Gamma) \tau}
\sin(\omega_q \tau).
\end{equation}
\begin{figure}[h] 
    \centering
    \includegraphics[width=1.0\linewidth]{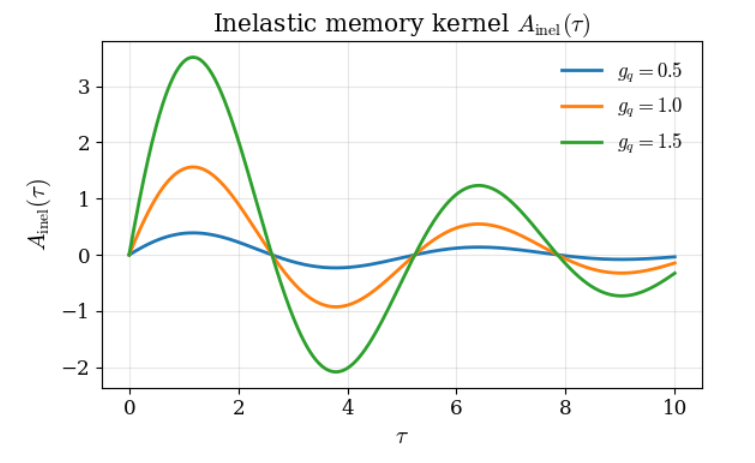}
    \caption{SCBA: $A_{inel}(\tau)$ versus $\tau$ Plot. }
\end{figure}\\
The inelastic contribution to the electronic spectral function acquires a finite temporal support determined by the phonon correlation time and the electron lifetime. As shown above, $A_{inel}(\tau)$ decays exponentially with a rate $\gamma_{q}+\Gamma$, where $\gamma_{q}$ originates from phonon damping and $\Gamma$ from electron self-consistency, while oscillating at the phonon frequency $\omega_{q}$. This explicitly demonstrates the emergence of non-Markovian memory in the inelastic scattering channel.\\

Diagrammatically, both approximation correspond to the same one-loop electron-boson diagram. The Born approximation(\ref{A},\ref{B})evaluates the loop using the bare propagator $G_{0}$, while SCBA replaces it with the dressed Green's function $G$, thereby incorporating lifetime feedback effects\cite{HaugHJauho}. This distinction becomes essential for inelastic processes due to the finite temporal correlations of the bosonic bath .\\    
\begin{figure}[h] 
\centering
\begin{tikzpicture}
\begin{feynman}
  \vertex (a);
  \vertex [right=2cm of a] (b);
  \diagram* {
    (a) -- [fermion] (b),
    (a) -- [scalar, half left, looseness=1.3] (b),
  };
\end{feynman}
\end{tikzpicture}
\caption{Lowest-order (Born) self-energy diagram for electron--phonon interaction.
The straight line denotes the bare electron Green's function \(G_0\), and the wavy line denotes the phonon propagator \(D\).}
\label{fig:born}
\end{figure}
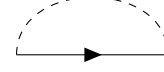
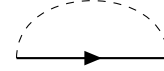
\begin{figure}[h] 
\centering
\begin{tikzpicture}
\begin{feynman}
  \vertex (a);
  \vertex [right=2cm of a] (b);
  \diagram* {
    (a) -- [fermion, thick] (b),
    (a) -- [scalar, half left, looseness=1.3] (b),
  };
\end{feynman}
\end{tikzpicture}
\caption{Self-consistent Born approximation (SCBA) self-energy diagram.
The thick fermion line denotes the fully dressed Green's function \(G\), while the phonon propagator remains bare.}
\label{fig:scba}
\end{figure}
 \begin{widetext}
\begin{minipage}{\linewidth}
\begin{figure}[H]
    \centering
    \includegraphics[width=1.0\linewidth]{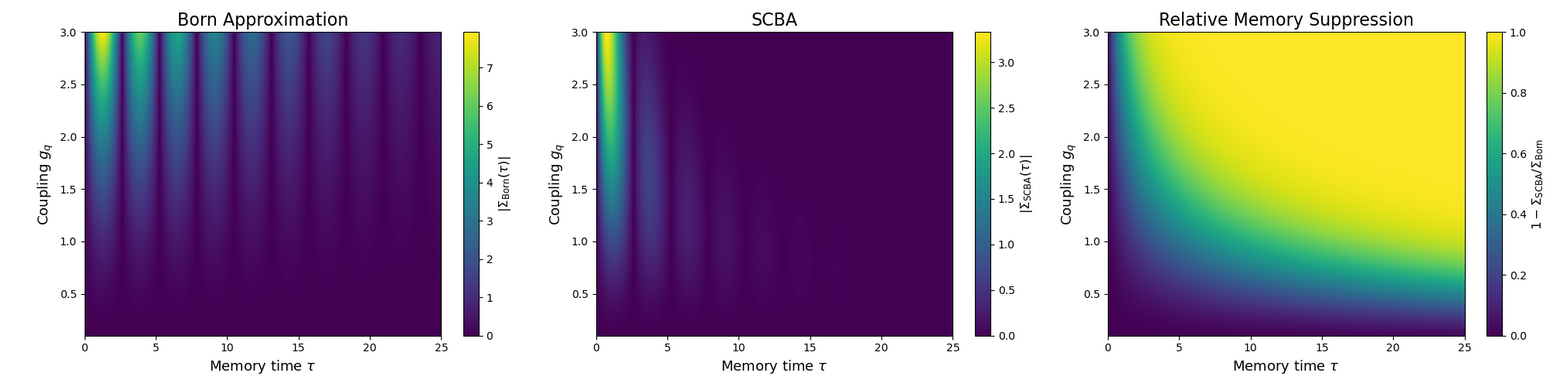}
    \caption{Heatmap comparison of the inelastic self-energy memory kernel within the Born approximation and self-consistent Born approximation (SCBA) as functions of coupling strength $g_q$ and memory time $\tau$. The Born kernel retains long-lived oscillatory temporal correlations over a broad coupling range, reflecting strong non-Markovian memory inherited from the phonon bath. In contrast, SCBA introduces self-consistent quasiparticle damping that progressively localizes the memory kernel near $\tau=0$, thereby suppressing long-time temporal correlations. The right panel shows the relative memory suppression 
$
R = 1-\frac{\Sigma_{\rm SCBA}}{\Sigma_{\rm Born}}$,
where $R\approx0$ indicates that Born and SCBA behave similarly, while $R\approx1$ signifies strong suppression of memory by self-consistent scattering, illustrating the crossover toward increasingly Markovian dynamics at stronger coupling.}
    \label{flowchart}
\end{figure}

\end{minipage}
 \end{widetext}
The discussion so far has focused on understanding memory effects in quantum transport from a nonequilibrium quantum field-theoretical perspective. By revisiting the Schwinger-keldysh formalism and the Kadanoff-Baym equations, we identified how elastic and inelastic scattering generate qualitatively different temporal structures in Green's functions, with time-local self-energies leading to a rapidly decaying correlations and interaction-induced self-energies giving rise to extended temporal memory. This analysis clarifies the microscopic origin of memory effects and establishes the theoretical foundation from which the nonequilibrium Green's function(NEGF) formalism naturally emerges.\\ Apart from that we present spectral functions for increasing coupling strengths within both Born and SCBA to
directly probe the evolution of temporal memory. While the Born approximation yields weak broadening and persistent long-time correlations, SCBA introduces lifetime effects through repeated scattering, producing stronger damping of the propagator. Consequently, increasing $g_q$ progressively suppresses long-time tails and weakens non-Markovian behavior. The comparison therefore reveals how
interaction-induced broadening drives the system toward effectively Markovian dynamics.\\
\section{Renormalization-Group Interpretation of Markovian and Non-Markovian Dynamics}\label{Fifth}
\label{sec:RG_memory}
The distinction between Markovian and non-Markovian dynamics can be understood within a renormalization-group (RG) framework \cite{meirinhos2022adaptive}, where fast degrees of freedom are integrated out to obtain an effective theory for slow variables. Starting from the full partition function,
\begin{equation}
Z = \int \mathcal{D}\phi\, e^{iS[\phi]},
\end{equation}
coarse-graining \cite{Delamotte:2007pf,meyer2019nonmarkovianoutofequilibriumdynamicsgeneral}leads to an effective action $S_{\text{eff}}$ that encodes the temporal structure of the dynamics.

In the absence of environmental coupling, the effective action remains local in time, resulting in equations of motion that depend only on the instantaneous state, characteristic of Markovian dynamics. By contrast, coupling to additional degrees of freedom (e.g., a phonon bath) and integrating them out generates temporally nonlocal terms.which introduce memory effects.

Consequently, the dynamics is governed by a Dyson equation with a time-convolution kernel(equation \ref{eq3}).
where the self-energy $\Sigma(t,t') \sim D(t-t')G(t,t')$ encodes the memory. Thus, non-Markovian behavior directly arises from the temporal nonlocality of the effective action induced by coarse-graining. A detailed derivation is provided in Appendix \ref{Appendix C}.

\section{Memory Signatures from conventional NEGF}\label{Sixth}
Apart from getting Markovian and non-Markovian signatures through temporal correlations we will see that  by analyzing transmission function from conventional NEGF \cite{KeremCamsari,Thakur,Datta_1995},  how it acts as a diagnostic tool for memory and hence Markovian versus non-Markovian signatures in the diagrammatical level. \\
In steady state limit, however, time-translational invariance is restored. All two-time quantities then depend on the relative time difference,
 \begin{equation}
     G(t,t')\to G(t-t'), \qquad \Sigma(t,t')\to \Sigma(t-t').
 \end{equation}
 This allows a Fourier transformation with respect to the relative time, yielding the conventional energy- domain NEGF formalism:
 \begin{equation}
     G(E)= \int d(t-t')e^{iE(t-t')G(t-t')}.
 \end{equation}
 In the energy representation, the Dyson equation becomes algebric\cite{yanik,Thakur},
 \begin{equation}
     G^R(E)=[E-H_{0}-\Sigma^R(E)]^{-1},
 \end{equation}
 where the self-energy $\Sigma^R(E)$ retains the dynamical information of the bath. Importantly, temporal memory encoded in the nonlocal kernel $\Sigma(t-t')$ manifests in energy space as a nontrivial frequency dependency of $\Sigma(E)$. Thus, non-Markovian effects in time corresponds to energy-dependent broadening and renormalization in the steady-state spectral properties.\\
 Now, for elastic scattering, the self-energy is taken to be local in energy and proportional to the Green’s function evaluated at the same energy \cite{Pragya}:
 \begin{equation}
     \Sigma^{el}_{S}(E)=DG^R(E).
 \end{equation}
 And the corresponding Green's function is:
 \begin{equation}
    G(E) =\left[(E+i\eta)I-H_0-\Sigma_{L}(E)-\Sigma_{R}(E)-D_{el}G(E)\right]^{-1}
    \end{equation}
Here D denotes the effective scattering strength. Since the self-energy depends on the Green’s function itself, the Dyson equation must be solved self-consistently. This form represents energy-local renormalization and corresponds to an effectively Markovian description in steady state.
For inelastic scattering, the self-energy mixes different energy sectors through processes involving energy exchange $\pm\Delta$ \cite{Pragya}:
\begin{equation}
    \Sigma^{inel}_{s}(E)=D[G^R(E+\Delta)+G^R(E-\Delta)]
\end{equation}
The corresponding GF becomes\cite{Pragya}
\begin{multline}
    G(E) =[(E+i\eta)I-H_0-\Sigma_{L}(E)\\
    -\Sigma_{R}(E)-D_{+} G(E - \Delta) + D_{-} G(E + \Delta)]^{-1}
\end{multline}
This explicitly introduces energy dependence beyond simple local broadening. In the time domain, such energy mixing corresponds to a temporally nonlocal (retarded) memory kernel. Upon Fourier transformation to steady state, this memory manifests as a frequency-dependent self-energy.\\
 And the transmission function\cite{PhysRevLett.68.2,Datta_1995,Datta_2005} within this standard NEGF formalism is given by
 \begin{equation}
     T(E)=Tr[\Gamma_{L}(E)G^R(E)\Gamma_{R}(E)G^A(E)],
 \end{equation}
 with \begin{equation}
     \Gamma_{\alpha}(E)= i[\Sigma^R_{\alpha}(E)-\Sigma^A_{\alpha}(E)].
 \end{equation}
 Hence, any memory-induced structure in $\Sigma(E)$ directly translates into measurable features in $T(E)$, providing a steady-state signature of the underlying temporal correlations.
\begin{center}
    \textbf{MODEL HAMILTONIANS}\\
\end{center}
We consider two representative two-dimensional systems: the Hofstadter model\cite{Hofstadter} and a long-range RKKY Hamiltonian\cite{RKKYrange,yanik}. The Hofstadter model exhibits a highly nontrivial band structure due to magnetic flux, while the RKKY model introduces long-range oscillatory interactions. This choice allows us to demonstrate that the emergence of Markovian versus non-Markovian transport behavior is independent of the underlying spectral complexity or interaction range, highlighting the generality of our results.
\subsubsection{\textbf{Two-Dimensional Hofstadter Hamiltonian}}
To investigate quantum transport in the presence of magnetic flux, we consider the two-dimensional Hofstadter Hamiltonian model defined on square lattice. The Hamiltonian\cite{Hofstadter} is
\begin{equation}
    H_{Hof}=-t\sum_{m,n}(e^{i2\pi\alpha n}c^\dagger_{m+1,n}c_{m,n}+c^\dagger_{m,n+1}c_{m,n}+h.c.)
\end{equation}
where t is nearest-neighbor hopping amplitude, $c^{\dagger}_{m,n}(c_{m,n})$ creates(annihilates) an electron at lattice site (m,n) and $\alpha=\frac{\phi}{\phi_{0}}$ is the magnetic flux per plaquette in units of the flux quantum $\phi_{0}=\frac{h}{e}$. The complex phase factor arises from the Peierls substitution, which incorporates the effect of a perpendicular magnetic field $B=B\hat{z}$. In the Landau gauge $A=(0,B_{x},0)$, hopping along one lattice direction acquires a phase 
\begin{center}
    $t\to t\times exp(i\frac{e}{h}\int A.dl)$.
\end{center}
This breaks ordinary translational invariance and leads to magnetic bloch bands.\\
In our transport this Hamiltonian serves as a prototype of a magnetically frustrated 2D lattice systems with complex hopping phases.
\subsubsection{\textbf{Two-Dimensional RKKY Hamiltonian}}
To explore long-range correlated interactions, we consider a 2D Ruderman–Kittel–Kasuya–Yosida (RKKY) model\cite{yanik,RKKYrange} defined as
\begin{equation}
    H_{RKKY}=\sum_{i\ne j}J(r_{i}-r_{j})S_{i}.S_{j},
\end{equation}
where $S_i$ denotes the localized spin at lattice site i, J(r) is the distance-dependent exchange interaction and $r_{i}$ is the position of lattice site i.\\
\textbf{ RKKY coupling in 2D :  }\\
In two dimensions, the exchange interaction mediated by itinerant electrons takes the asymptotic form
\begin{center}
  $J(r)=J_{0} \frac{cos(2K_{F}r)}{r^2}$ \cite{RKKYrange},
\end{center}
where $k_{F}$ is the fermi wavevector and $r=|r_{i}-r_j|$.\\
\begin{center}
\textbf{STEADY-STATE SIGNATURES OF MARKOVIAN AND NON-MARKOVIAN SCATTERING}
\end{center}
Any memory-induced modifications in the spectral function, such as peak broadening, shifts, or the emergence of additional structure due to energy-dependent self-energies are directly reflected in the transmission function. This establishes T(E) as an experimentally accessible probe of memory effects, justifying its use to identify Markovian and non-Markovian signatures\ref{Appendix D}.
 So,to identify steady-state manifestation of memory effects, we analyze the transmission $T(E)$ as a function of the effective scattering mechanisms for both the Hofstadter and RKKY Hamiltonians under elastic and inelastic scattering mechanisms. Where,

\begin{equation}
    \lambda=||Im\Sigma_{S}(E)||
\end{equation}
is effective scattering strength.
\begin{figure} [h] 
    \centering
    \includegraphics[width=1.0\linewidth]{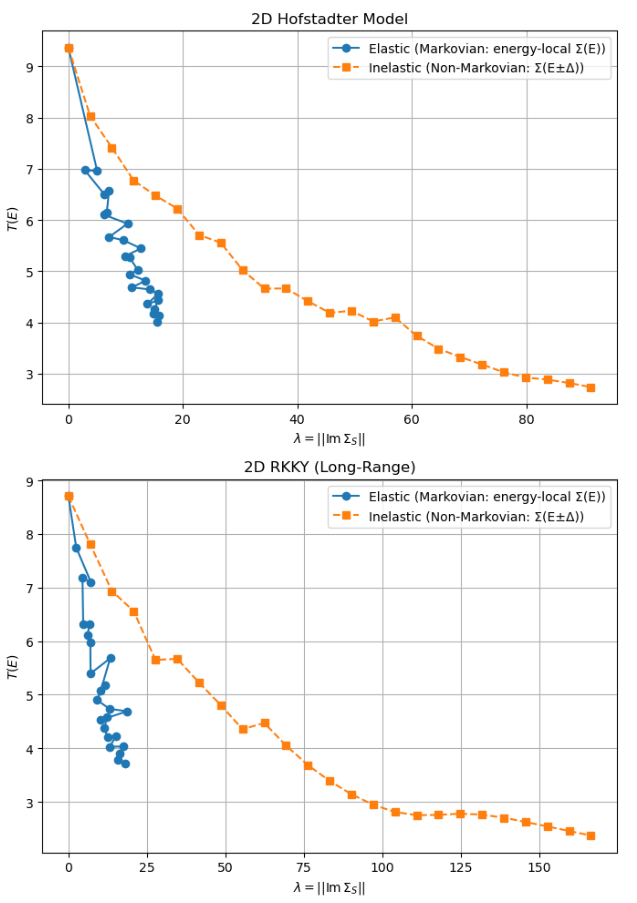}
    \caption{Transmission T(E) versus effective scattering strength $\lambda=||Im\Sigma_{s}||$ for the 2D Hofstadter(top) and 2D RKKY(bottom) models.Numerical scheme for these plot is in \ref{Appendix D}.}
\end{figure}
\subsubsection{\textbf{Elastic Scattering: Markovain-like Behavior}}
In the elastic case, the scattering self-energy\cite{KeremCamsari} is local in energy and proportional to the Green’s function evaluated at the same energy. Consequently, the self-energy acts primarily as an energy-local broadening mechanism without inducing energy mixing. This corresponds to an effectively Markovian description in which the temporal self-energy kernel is short-ranged, leading in steady state to a weakly structured, nearly monotonic renormalization of transmission.
For both the Hofstadter and RKKY systems, the elastic transmission curves exhibit smooth suppression with increasing effective broadening. The absence of additional structure indicates that scattering primarily induces local decoherence without generating dynamical correlations across different energy sectors. Thus, elastic scattering behaves as a steady-state analogue of Markovian dissipation.
\subsubsection{\textbf{Inelastic scattering: Non-Markovian Signatures}}
In contrast, the inelastic self-energy explicitly mixes different energies through terms of the form $G(E\pm\Delta)$ generating a hierarchy of coupled Dyson equations. In the time domain, such energy mixing corresponds to a temporally nonlocal (retarded) kernel, i.e., genuine memory. Upon Fourier transformation to steady state, this temporal nonlocality manifests as a frequency-dependent self-energy.
This energy dependence modifies the transmission beyond simple broadening, producing a smoother yet nontrivial renormalization profile. The response is no longer governed solely by local decoherence at E, but instead reflects spectral redistribution induced by energy exchange processes. Such behavior represents the steady-state fingerprint of non-Markovian dynamics.
\subsubsection{\textbf{Universality Across Hofstadter and RKKY Hamiltonians}}
Importantly, these qualitative distinctions persist for both the Hofstadter lattice and the long-range RKKY system. Although the underlying band structures differ substantially topological magnetic minibands \cite{Hofstadter1976} in the Hofstadter case and long-range oscillatory coupling in the RKKY case, the contrast between elastic (energy-local) and inelastic (energy-mixing) self-energies remains robust. This demonstrates that the observed Markovian versus non-Markovian signatures originate from the structure of the scattering kernel rather than from specific lattice details.
\section{CONCLUSION}
 This work develops a coherent framework for understanding Markovian and non-Markovian dynamics in quantum transport by linking microscopic theory, diagrammatic structure,
and renormalization-group interpretation. The analysis of spectral functions in Section \ref{Third} demonstrates that elastic processes predominantly exhibit Markovian signatures characterized by simple
exponential decay, whereas inelastic scattering introduces oscillatory and long-lived structures that encode non-Markovian memory effects. Section \ref{Fourth} further strengthens this picture through a diagrammatic comparison between the Born and self-consistent Born approximations, showing that non-Markovian behavior becomes increasingly pronounced with stronger coupling strength $g_q$, under-scoring the importance of self-consistency in capturing temporal correlations. From a coarse graining perspective, the distinction between 1PI and 2PI \cite{meirinhos2022adaptive} effective actions in Section \ref{Fifth} clarifies that stronger coarse graining (1PI) naturally suppresses memory and favors Markovian dynamics, while weaker coarse graining (2PI) retains temporal nonlocality and memory kernels. Finally, Section \ref{Sixth} demonstrates that these memory effects manifest directly in observable transport quantities, where signatures of non-Markovian dynamics appear in the transmission function of two paradigmatic two-dimensional model Hamiltonians, thereby connecting formal theoretical structure with measurable transport behavior.\\
Existing studies have primarily addressed either driven transport within the NEGF framework or non-Markovian dynamics in open quantum systems. However, the role of external driving in controlling memory effects, and their manifestation in steady-state transport quantities, has not yet been systematically explored. So, that can be addressed in my future work.\\

\textbf{Acknowledgements:} The author wants to acknowledge and thank Dr. Niladri Sarkar, Mr. Sayan Das and Dr. Sourav Dey for very useful discussions. Also P.C. wants to acknowledge BITS-PILANI for support.

\nocite{*}
\bibliographystyle{unsrturl}
\bibliography{name}

\section{Appendix (A)}\label{A}
\subsection{construction of Kadanoff-Baym equation}
Begin from the exact operator identity on the Keldysh contour\cite{NEGfNEW} 
\begin{equation}
G(1,2)= G_{0}(1,2)+\int_{c}d3d4G_{0}(1,3)\Sigma(3,4)G(4,2)
\end{equation}
Now multiply both sides by $G^{-1}(1,\bar{1})$ and integrate over $\bar{1}$
\begin{multline}
    \int_{c}d\bar{1}\,G_{0}^{-1}(1,\bar{1})G(\bar{1},2)=\int_{c}d\bar{1}\, G_{0}^{-1}(1,\bar{1}) G_{0}(\bar{1},2)\\  +\int_{c}d\bar{1} \, d3\,d4\,G_{0}^{-1}(1,\bar{1})G(\bar{1},3)\,\Sigma(3,4)\,G(4,2)
\end{multline}

we know that
\begin{equation}
    \int_{C}d\bar{1}G_{0}^{-1}(1,\bar{1})G_{0}(\bar{1},2)=\delta_{c}
(1-2)
\end{equation}
Similarly,
\begin{equation}
    \int_{C}d\bar{1}G_{0}^{-1}(1,\bar{1})G_{0}(\bar{1},3)=\delta_{c}
(1-3)
\end{equation}
So
\begin{equation}
    \int_{c}d3d4\delta(1-3)\Sigma(3,4)G(4,2)=\int_{c}d4\Sigma(1,4)G(4,2)
\end{equation}
So we obtain
\begin{equation}
    \int_{C}d3G_{0}^{-1}(1,3)G(3,2)= \delta_{c}(1-2)+\int_{c}d3\Sigma(1,3)G(3,2)
    \label{eq:Dyson}
\end{equation}
The contour ordered Green's function
\begin{equation}
    G_{0}(1,2) = -i\langle T_{\mathcal C}\psi(1)\psi^\dagger(2)\rangle
\end{equation}
Now using free equation of motion of the field and Free Hamiltonian is quadratic
\begin{equation}
    H_0 = \int d\mathbf{x}\;
\psi^\dagger(\mathbf{x})\,h\,\psi(\mathbf{x}),
\end{equation}
and the Heisenberg equation of motion gives,
\begin{equation}
i\partial_{t} \psi(1)
=
h(1)\psi(1).
\end{equation}
This is free Schrodinger's equation so
\begin{equation}
    (i\partial_{t}-h)\psi=0
\end{equation}
 Applying it to the first argument
 \begin{equation}
     (i\partial_{t}-h(1))G_{0}(1,2)=-i\langle{T_{c}}(i\partial_{t}-h(1))\psi(1)\psi(2)\rangle
 \end{equation}
 that produces a delta function
 \begin{equation}
(i\partial_{t}-h(1))G_{0}(1,2)=\delta_{c}(1-2)
 \end{equation}
 and
 \begin{equation}
     \int_{c}d3G_{0}^{-1}(1,3)G_{0}(3,2)= \delta_{c}(1-2)
 \end{equation}
 Comparing with the differential equation we get,
 \begin{equation}
     G_{0}^{-1}(1,2)=(i\partial_{t}-h(1))\delta_{c}(1-2)
 \end{equation}
 Similarly,
 \begin{equation}
     G_{0}^{-1}(1,3)=(i\partial_{t}-h(1))\delta_{c}(1-3)
 \end{equation}
By inserting this in eq. \eqref{eq:Dyson} 
\begin{multline}
    \int_{c}d3(i\partial_{t}-h(1))\delta_{c}(1-3)\,G(3,2)=\delta_{c}(1-2)\\+\int_{c}d3\,\Sigma(1,3)\,G(3,2)
\end{multline}

 Using the delta function
     \begin{equation}
         (i\partial_{t}-h(1))G(1,2)=\delta_{c}(1-2)+\int_{c}d3\Sigma(1,3)G(3,2)
     \end{equation}
 This is Schwinger-Dyson equation on contour/Kadanoff-Baym equation.
\section{Appendix (B)}\label{B}
\subsection{Born Approximation for the Electron--Phonon Self-Energy}
We consider the electron--phonon interaction Hamiltonian
\begin{equation}
H_{\text{inel}}(t)
=
\sum_{i,q} g_q\,
c_i^\dagger(t)c_i(t)
\big(b_q(t)+b_q^\dagger(t)\big),
\end{equation}
where $b_q^\dagger$ and $b_q$ create and annihilate phonons of mode $q$.

The interacting contour-ordered Green's function is defined as \cite{ochoa2025semiclassical}
\begin{equation}
G(1,2)
=
-\,i\,
\frac{
\left\langle
T_{\mathcal C}\,
\psi(1)\psi^\dagger(2)\,
\exp\!\left[-i\int_{\mathcal C} dt\, H_{\text{inel}}(t)\right]
\right\rangle
}{
\left\langle
T_{\mathcal C}\,
\exp\!\left[-i\int_{\mathcal C} dt\, H_{\text{inel}}(t)\right]
\right\rangle
},
\end{equation}
where $T_{\mathcal C}$ denotes contour ordering.

\subsubsection{Second-Order Expansion}

The lowest non-vanishing contribution from the interaction arises at second order in $g_q$ (Born approximation),
\begin{multline}
\exp\!\left[-i\int_{\mathcal C} dt\, H_{\text{inel}}(t)\right]
\approx
1
-
\frac{1}{2}
\int_{\mathcal C} dt_3 dt_4\,\\
T_{\mathcal C}
H_{\text{inel}}(t_3)H_{\text{inel}}(t_4).
\end{multline}
Substituting this into the definition of $G(1,2)$ yields the second-order correction
\begin{multline}
\delta G(1,2)
=
(-i)^2
\int_{\mathcal C} dt_3 dt_4\;
\Big\langle
T_{\mathcal C}\,
\psi(1)\psi^\dagger(2)\,\\
H_{\text{inel}}(t_3)H_{\text{inel}}(t_4)
\Big\rangle_{\text{conn}},
\end{multline}
where only connected contributions are retained.

\subsubsection{Insertion of the Interaction Hamiltonian}

Inserting the explicit form of $H_{\text{inel}}$ gives
\begin{multline}
\delta G(1,2)
=
\sum_q |g_q|^2
\int_{\mathcal C} dt_3 dt_4\;
\Big\langle
T_{\mathcal C}\,
\psi(1)\psi^\dagger(2)\\
\big[\psi^\dagger\psi(b_q+b_q^\dagger)\big]_{t_3}
\big[\psi^\dagger\psi(b_q+b_q^\dagger)\big]_{t_4}
\Big\rangle .
\end{multline}

\subsubsection{Wick Contractions}

Since the phonons are harmonic and the initial density matrix factorizes into electronic and phononic parts, Wick's theorem applies separately to electrons and phonons.

The phonon contraction yields the phonon Green's function
\begin{equation}
\big\langle
T_{\mathcal C}\,
(b_q+b_q^\dagger)_{t_3}
(b_q+b_q^\dagger)_{t_4}
\big\rangle
=
-\,i\,D_q(3,4),
\end{equation}
where $D_q$ is the contour-ordered phonon propagator.

The connected fermionic contraction contributing to the self-energy gives
\begin{multline}
\big\langle
T_{\mathcal C}\,
\psi(1)\,
\psi^\dagger(t_3)\psi(t_3)\,
\psi^\dagger(t_4)\psi(t_4)\,
\psi^\dagger(2)
\big\rangle
\;\Rightarrow\;\\
G(1,4)\,G(3,2).
\end{multline}

\subsubsection{Identification of the Self-Energy}

Collecting all contributions, the second-order correction to the Green's function takes the form
\begin{equation}
\delta G(1,2)
=
\int_{\mathcal C} d3\, d4\;
G(1,3)\,
\Sigma(3,4)\,
G(4,2),
\end{equation}
which allows us to identify the electron--phonon self-energy as
\begin{equation}
\Sigma(3,4)
=
i\sum_q |g_q|^2\,
G(3,4)\,
D_q(3,4).
\end{equation}
\section{\textbf{APPENDIX C}}\label{Appendix C}
\subsection{Coarse-graining and effective theories}
The renormalization group (RG) provides a systematic procedure for deriving
effective low-energy or long-time dynamics by eliminating fast microscopic
degrees of freedom.
Starting from the full partition function
\begin{equation}
Z = \int \mathcal{D}\phi \, e^{iS[\phi]},
\end{equation}
By decomposing the field into slow and fast components with respect to short time(high-frequency fluctuations) and long time(low-frequency dynamics)
\begin{equation}
    \phi = \phi_{\text{slow}} + \phi_{\text{fast}},
\end{equation}
and integrate out the fast fluctuations,\\
\begin{multline}
Z
=
 \int \mathcal{D}\phi_{\text{slow}}\,
\left(
\int \mathcal{D}\phi_{\text{fast}}\,
e^{iS[\phi_{\text{slow}},\phi_{\text{fast}}]}
\right)\\
=
\int \mathcal{D}\phi_{\text{slow}}
e^{iS_{\text{eff}}[\phi_{\text{slow}}]}.
\end{multline}
The effective action $S_{\text{eff}}$ here, describes only the slow
degrees of freedom. The elimination of the microscopic information constitutes \emph{coarse-graining}.
Depending on which variables are retained after coarse-graining,
qualitatively different dynamical structures arise.
In particular, the distinction between the one-particle-irreducible (1PI) and two-particle-irreducible (2PI) effective actions\cite{Berges_2004,meirinhos2022adaptive,blaizot2021functional} can be interpreted as different levels of temporal coarse-graining, which directly determines
whether the resulting dynamics is Markovian or non-Markovian.

\subsection{Local dynamics and 1PI effective action (electronic sector)}

Let's start by considering the tight-binding Hamiltonian already discussed in \ref{eq7}
\begin{equation}
H_{0}=\sum_{i,j}t_{ij}c_i^\dagger c_j+\sum_i U_i c_i^\dagger c_i .
\end{equation}

Here $c_i^\dagger$ and $c_i$ denote fermionic creation and annihilation operators at lattice site $i$, $t_{ij}$ represents the hopping amplitude between sites $i$ and $j$, and $U_i$ denotes the on-site potential. Defining
\begin{equation}
h_{ij}=t_{ij}+U_i\delta_{ij},
\end{equation}
the Hamiltonian can be written in compact form as
\begin{equation}
H_0=\sum_{i,j}c_i^\dagger h_{ij}c_j .
\end{equation}

To analyze the dynamics we employ a path-integral representation using Grassmann fields $\psi_i(t)$ and $\bar{\psi}_i(t)$. The electronic action becomes
\begin{equation}
S[\bar\psi,\psi]
=
\int dt
\sum_{i,j}
\bar{\psi}_i(t)
\left[i\partial_t\delta_{ij}-h_{ij}\right]
\psi_j(t).
\end{equation}

The generating functional\cite{PhysRevB.102.205131} is
\begin{equation}
Z=\int D[\bar\psi,\psi]\,
e^{iS[\bar\psi,\psi]}.
\end{equation}

In order to construct the one-particle-irreducible (1PI) effective action, we introduce linear sources $J_i(t)$ and $\bar{J}_i(t)$,
\begin{equation}
Z[J,\bar J]=
\int D[\bar\psi,\psi]
\exp
\left[
iS[\bar\psi,\psi]
+
i\int dt(\bar J_i\psi_i+\bar\psi_i J_i)
\right].
\end{equation}

The connected generating functional\cite{meirinhos2022adaptive,blaizot2021functional} is defined as
\begin{equation}
W[J,\bar J]=-i\ln Z[J,\bar J].
\end{equation}

The classical fields are obtained from
\begin{equation}
\psi_i(t)=\frac{\delta W}{\delta \bar J_i(t)},
\qquad
\bar\psi_i(t)=-\frac{\delta W}{\delta J_i(t)}.
\end{equation}

The 1PI effective action is then obtained through the Legendre transform\cite{blaizot2021functional,meirinhos2022adaptive}
\begin{equation}
\Gamma[\bar\psi,\psi]
=
W[J,\bar J]
-
\int dt
(\bar J_i\psi_i+\bar\psi_i J_i).
\end{equation}

For the quadratic electronic action considered here, the path integral is Gaussian and the effective action reduces to
\begin{equation}
\Gamma[\bar\psi,\psi]
=
\int dt
\sum_{i,j}
\bar\psi_i(t)
\left[i\partial_t\delta_{ij}-h_{ij}\right]
\psi_j(t).
\end{equation}

The physical dynamics follows from the stationary condition
\begin{equation}
\frac{\delta\Gamma}{\delta\bar\psi_i(t)}=0,
\end{equation}
which yields the equation of motion
\begin{equation}
i\partial_t\psi_i(t)=\sum_j h_{ij}\psi_j(t).
\end{equation}

Since the evolution equation depends only on the instantaneous value of the field at time $t$, the dynamics is \textit{local in time} and therefore Markovian. Physically, this reflects the fact that the Hamiltonian contains only coherent electronic hopping and does not include any environmental degrees of freedom that could generate memory effects.

\subsection{Phonon bath and 2PI effective action: emergence of non-Markovian dynamics}

To incorporate inelastic scattering processes and memory effects, we now couple the electronic system to a phonon bath. The total Hamiltonian becomes
\begin{equation}
H=H_0+H_{ph}+H_{e-ph}.
\end{equation}

The phonon bath is described by
\begin{equation}
H_{ph}=\sum_q \omega_q b_q^\dagger b_q,
\end{equation}
while the electron--phonon interaction is taken in the Holstein form
\begin{equation}
H_{e-ph}=\sum_{i,q} g_q\, c_i^\dagger c_i(b_q+b_q^\dagger).
\end{equation}

Introducing a bosonic field $\phi_q(t)$, the phonon action reads
\begin{equation}
S_{ph}
=
\frac12
\int dt
\sum_q
[\dot{\phi}_q^2-\omega_q^2\phi_q^2].
\end{equation}

The interaction term becomes
\begin{equation}
S_{e-ph}
=
\int dt
\sum_{i,q}
g_q\,\bar\psi_i(t)\psi_i(t)\phi_q(t).
\end{equation}

The full generating functional\cite{Berges_2004,meirinhos2022adaptive} is therefore 
\begin{equation}
Z=\int D[\bar\psi,\psi,\phi]\,
e^{i(S_e+S_{ph}+S_{e-ph})}.
\end{equation}

Since the phonon action is quadratic, the phonon fields can be integrated out exactly, leading to an effective electronic action\cite{PhysRevB.102.205131}
\begin{equation}
Z=\int D[\bar\psi,\psi]\,
e^{iS_{\text{eff}}}.
\end{equation}

The resulting interaction term is
\begin{equation}
S_{int}
=
-\frac12
\int dt\,dt'
\,n_i(t)D_{ij}(t-t')n_j(t'),
\end{equation}
where
\begin{equation}
n_i(t)=\bar\psi_i(t)\psi_i(t)
\end{equation}
and
\begin{equation}
D_{ij}(t-t')
=
\sum_q g_q^2 D_q(t-t')
\end{equation}
is the phonon propagator.

The interaction kernel $D(t-t')$ couples electronic densities at two different times, indicating that the effective dynamics is inherently nonlocal in time.

To treat such temporal correlations, it is convenient to employ the two-particle-irreducible (2PI) effective action formalism. Introducing the contour-ordered Green function
\begin{equation}
G_{ij}(t,t')=-i\langle T\psi_i(t)\psi_j^\dagger(t')\rangle,
\end{equation}
the 2PI effective action takes the form\cite{meirinhos2022adaptive,blaizot2021functional}
\begin{equation}
\Gamma[\psi,G]
=
S_{\text{eff}}[\psi]
+
\frac{i}{2}\mathrm{Tr}\ln G^{-1}
+
\frac{i}{2}\mathrm{Tr}(G_0^{-1}G)
+
\Gamma_2[\psi,G],
\end{equation}
where $G_0^{-1}=(i\partial_t-h)\delta(t-t')$.

The stationarity condition
\begin{equation}
\frac{\delta\Gamma}{\delta G}=0
\end{equation}
leads to the Dyson equation\cite{meirinhos2022adaptive,blaizot2021functional}
\begin{equation}
G^{-1}=G_0^{-1}-\Sigma,
\end{equation}
with the electron--phonon self-energy
\begin{equation}
\Sigma_{ij}(t,t')=i\,\delta_{ij}\,D(t-t')G_{ii}(t,t').
\end{equation}

The corresponding equation of motion reads\cite{meirinhos2022adaptive}
\begin{equation}
(i\partial_t-h)G(t,t')
-
\int dt_1
\Sigma(t,t_1)G(t_1,t')
=
\delta(t-t').
\end{equation}

Unlike the local equation obtained in the 1PI formulation, this equation contains a time convolution over the past history of the system, reflecting the retarded interaction mediated by the phonon bath. Consequently, the dynamics becomes non-Markovian, with memory effects encoded in the phonon propagator $D(t-t')$.

\section{APPENDIX (D)}\label{Appendix D}
\begin{center}
{\textbf{[1].Numerical Scheme for Quantum Transport Calculations}}
\end{center}
\label{appendix:negf_flowchart}

The calculation of the transport properties in this work follows the Non-Equilibrium Green's Function (NEGF) formalism, specialized for a device coupled to two semi-infinite leads. Figure \ref{flowchart} illustrates the iterative numerical procedure used to obtain the converged surface Green's functions and subsequent transport observables.

The process begins with an iterative calculation of the surface Green's functions $g_1$ and $g_2$ for the left and right leads, respectively. We utilize the decimation technique (or self-consistent iteration) where:
\begin{equation}
    g^{-1} = (E + i\eta)I - \alpha - \beta g \beta^\dagger
\end{equation}
Here, $\alpha$ represents the Hamiltonian of the unit cell and $\beta$ represents the coupling between cells. Once the convergence criteria for $g_{1,2}$ are satisfied, we compute the self-energies $\Sigma_{1,2}$, which describe the electronic coupling between the leads and the central scattering region. 

From these, the broadening matrices $\Gamma_{1,2}$ are derived, allowing for the evaluation of the retarded Green's function $[G]$ of the device. Finally, the transmission $T(E)$, the spectral function $[A]$, and the correlation function $[G^n]$ are calculated. This framework allows us to systematically analyze the dependence of the transmission $T(E)$ on the imaginary part of the self-energy $\|\text{Im}\,\Sigma_S\|$, which represents the escape rate of carriers into the leads.
 \begin{widetext}
\begin{minipage}{\linewidth}
\begin{figure}[H]
    \centering
    \includegraphics[width=0.8\linewidth]{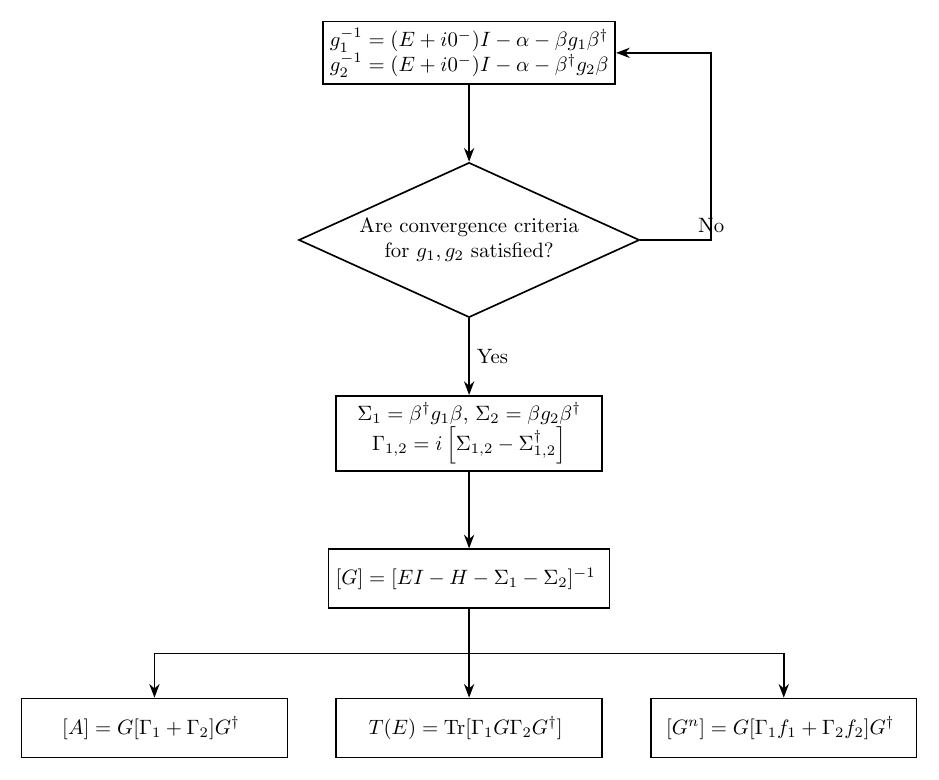}
    \caption{Flowchart to explain NEGF methodology \cite{Thakur}}
    \label{flowchart}
\end{figure}

\end{minipage}
 \end{widetext}

\begin{center}
{\textbf{[2].Explicit Relation Between Spectral Function and Transmission Function}}
\end{center}
\nopagebreak 
Within the conventional NEGF formalism, the spectral function \cite{Thakur}
\begin{equation}
A(E) = i\left[G^R(E) - G^A(E)\right]
\end{equation}
encodes the density of available states and carries direct signatures of the self-energy, including memory effects through its energy dependence. In particular, non-Markovian dynamics manifests as nontrivial structure in $A(E)$, arising from the energy dependence of the self-energy.

The transmission function,
\begin{equation}
T(E) = \mathrm{Tr}\left[\Gamma_L G^R(E)\,\Gamma_R G^A(E)\right],
\end{equation}
is directly related to the spectral properties of the system. For proportional coupling, it can be expressed as
\begin{equation}
T(E) \propto \mathrm{Tr}\left[\Gamma\, A(E)\right],
\end{equation}
indicating that transport is governed by the same spectral features encoded in $A(E)$.

Consequently, any memory-induced modifications in the spectral function such as peak broadening, shifts, or the emergence of additional structure due to energy-dependent self-energies are directly reflected in the transmission function. This establishes $T(E)$ as an experimentally accessible probe of memory effects, justifying its use to identify Markovian and non-Markovian signatures.

\end{document}